\documentclass[12pt]{article}

\usepackage{bbm}
\usepackage{epsfig}
\usepackage{array}
\usepackage{float}
\usepackage{dsfont}
\usepackage{amstext}
\usepackage{rotating}
\usepackage{a4}
\usepackage{a4wide}
\usepackage{cite}
\usepackage{amsmath} 

\def\be{\begin{equation}}
\def\ee{\end{equation}}
\def\gs{\mathrel{
   \rlap{\raise 0.511ex \hbox{$>$}}{\lower 0.511ex \hbox{$\sim$}}}}
\def\ls{\mathrel{
   \rlap{\raise 0.511ex \hbox{$<$}}{\lower 0.511ex \hbox{$\sim$}}}}

\newcommand{\ba}{\begin{array}{c}}
\newcommand{\baz}{\begin{array}{cc}}
\newcommand{\barrr}{\begin{array}{rrr}}
\newcommand{\bad}{\begin{array}{ccc}}
\newcommand{\bav}{\begin{array}{cccc}}
\newcommand{\baf}{\begin{array}{ccccc}}
\newcommand{\bea}{\begin{equation} \begin{array}{c}}
\newcommand{\eea}{ \end{array} \end{equation}}
\newcommand{\ea}{\end{array}}
\newcommand{\D}{\displaystyle}
\newcommand{\dms}{\mbox{$\Delta m^2_{\odot}$}}

\newcommand{\gsim}{\raise0.3ex\hbox{$\;>$\kern-0.75em\raise-1.1ex\hbox{
   $\sim\;$}}} 
\newcommand{\lsim}{\raise0.3ex\hbox{$\;<$\kern-0.75em\raise-1.1ex\hbox{
   $\sim\;$}}}

\begin{document}

\title{
\vskip 0.4cm
\bf \Large
Possible Alternatives to Tri-bimaximal Mixing}

\author{
Carl H.~Albright$^{a,b}$\thanks{email: 
\tt albright@fnal.gov}~\mbox{ 
},~~Alexander Dueck$^c$\thanks{email:
\tt alexander.dueck@mpi-hd.mpg.de}~\mbox{
},~~Werner Rodejohann$^c$\thanks{email: 
\tt werner.rodejohann@mpi-hd.mpg.de}
\\\\
{\normalsize \it$^a$Department of Physics, Northern Illinois University,}\\
{\normalsize \it DeKalb, Illinois 60115, USA}\\ \\ 
{\normalsize \it$^b$Fermi National Accelerator Laboratory,}\\
{\normalsize \it Batavia, Illinois 60510, USA}\\ \\ 
{\normalsize \it$^c$Max--Planck--Institut f\"ur Kernphysik,}\\
{\normalsize \it  Postfach 103980, D--69029 Heidelberg, Germany} 
}

\date{}
\maketitle
\thispagestyle{empty}
\vspace{0.8cm}
\begin{abstract}
\noindent  
Possible alternatives to tri-bimaximal mixing are presented based on  
other symmetry principles, and their predictions for $|U_{e3}|$, 
$\sin^2 \theta_{12}$ and $\sin^2 \theta_{23}$ 
are compared to the present
neutrino mixing data.  In some cases perturbations are required to give 
better agreement with the data, and the use of a minimal approach is
illustrated.  
Precise experimental determinations of the mixing 
parameters will be required to decipher the correct mixing pattern
and to pin down the appropriate flavor symmetry.

\end{abstract}

\newpage
\section{\label{sec:intro}Introduction}
The first discoveries of neutrino oscillations arose from observations of 
the depletions of atmospheric muon-neutrinos \cite{atm0} 
and solar electron-neutrinos \cite{sol0}, 
relative to their expected predictions.  In efforts to understand these 
findings, many theorists adopted top-down approaches in attempts to construct
models which would explain the data.  For this purpose, various forms of 
the neutrino and charged lepton mass matrices were postulated, some applied
directly to the light left-handed neutrino mass matrix, while other more
ambitious efforts invoked the seesaw mechanism partly 
using also the framework of grand unified models. 
Examples to constrain the mass matrices 
involved the assignment of texture zeros, the 
use of a vertical family symmetry group, and/or the selection of a 
horizontal flavor symmetry, usually of a continuous type such as $U(1),\ 
SU(2)$ or $SU(3)$.  The more complete models and their predictions differed by 
the choice of family and flavor symmetries, and the fermion and 
Higgs representation assignments made in the construction of the unknown 
Yukawa interactions needed to extend the Standard Model.  

As the oscillation data became more accurate with refinements in the 
atmospheric \cite{atm} and solar \cite{sol} neutrino experiments and 
introduction of land-based reactor \cite{rea} 
and long baseline neutrino \cite{lbl} experiments, bottom-up approaches to
construct models became more feasible.  Among the first to realize the 
mixing data were pointing to a rather simple construction were Harrison, 
Perkins and Scott \cite{HPS}, who coined the 
phrase ``tri-bimaximal mixing''.  In this
scheme the atmospheric neutrino mixing angle  $(\theta_{23})$ is maximal 
$45^\circ$, the reactor neutrino mixing angle $(\theta_{13})$ vanishes, while
the solar neutrino mixing angle is $\theta_{12} \simeq 35.3^\circ$, such 
that $\sin^2 \theta_{12} = \frac 13$.  
With this tri-bimaximal mixing (TBM) texture in mind, many models have been 
constructed based on the discrete symmetry groups such as 
$S_3,\ A_4,\ S_4,\ T'$, etc., with a vast majority using $A_4$ 
(see Refs.~\cite{AF,jap} for reviews on 
flavor symmetries, in particular $A_4,$ 
and Ref.~\cite{BR} for a classification of all existing 
(50+) type I seesaw, type II seesaw and non-seesaw $A_4$ models). 
While the tri-bimaximal mixing pattern lies within $1\sigma$ of the present
experimental fits, the best-fit points require some deviation from that 
pattern. 

It is fair to say that TBM dominates the theoretical 
literature in flavor model building\footnote{The original suggestion
of tri-bimaximal mixing was a purely phenomenological Ansatz and only 
later shown to be obtainable in dedicated flavor models.}. 
We remind the reader that attempts to explain the mixing data based 
on grand unified models using continuous flavor symmetry groups were also
reasonably successful in explaining the mixing data (see Ref.~\cite{AR2} 
for a list of 13 valid $SO(10)$ models in agreement with current data). 
This raises the issue whether there indeed exists some hidden flavor symmetry, 
such as $A_4$, or whether the nearly observed TBM mixing is accidental 
in nature.  Reference \cite{AR2} tried to attack this issue from the point 
of perturbing the neutrino mass matrix $m_\nu^{\rm TBM}$ corresponding to TBM. 
It was argued that when relative corrections to the mass matrix entries 
are applied, the value of $|U_{e3}|$ can be crucial to distinguish TBM 
from grand unified theories. 
A very recent paper \cite{AS} has shown that mass matrices which are 
significantly different from $m_\nu^{\rm TBM}$ are also allowed. 
It is thus important not to focus solely on one particular mixing
scheme, such as TBM, but to look for other options as well.  
In any case, it is apparent that very 
accurate experimental determinations of the neutrino mixing parameters will
be required in order to pin down the source of the flavor mixing. 

In the spirit of the above considerations, we point out 
in this letter the existence of a plethora of 
alternatives to TBM and explore a number of other possibilities for the 
neutrino mixing matrix. We wish to stress that many of the mixing
scenarios that we describe are allowed by the current data equally
well. Some of them have been obtained in models with the flavor symmetry
specified at the outset, and very often the choice of symmetry group is 
motivated by geometrical considerations. Good examples here are the two 
golden ratio possibilities for the solar neutrino mixing angle. 
Among the other examples we give is trimaximal mixing, where only the second 
column of the tri-bimaximal mixing matrix with equal flavor contributions is 
postulated.  Variations of this theme make the invariant assumption for the 
first or third column or one of the three rows. Yet another hypothesis 
involves quark-lepton complementarity where the quark and neutrino mixing 
matrices are related. Obviously, one should try to disentangle the huge number 
of proposed flavor models in order to sort out the correct one, or at least 
rule out many of the incorrect ones \cite{AC}. 

We should also mention that it is not unlikely that corrections to mixing
schemes may apply. Radiative corrections, effects of charged lepton
rotations, soft breaking, or ``NLO'' effects of the underlying 
flavor models are possibilities. The magnitude of the corrections 
relies heavily on the models which realize the respective scenarios, and
depend on a number of unknown parameters, such as neutrino masses or
CP phases. Let us mention, however, that radiative corrections are small 
for a normal hierarchy of neutrino masses, and that 
charged lepton rotations play no role if the 
symmetry basis coincides with the charged lepton mass basis. 
In principle one could perform for each scenario to be discussed in
the following a dedicated analysis of perturbations in
analogy, e.g., to the model-independent study for TBM in
Ref.~\cite{Ray}, or to studies for concrete models in Refs.~\cite{NLO}. 
In the present letter we neglect the study of these aspects, and
rather focus on pointing out the existence of a variety of 
alternatives to TBM, their possible physics motivation, and 
the ``unperturbed'' predictions of the scenarios. 
In principle, for each scenario considered, one can use the 
bottom-up approach to determine the neutrino mass matrix and presumably to 
construct a model based on some discrete flavor symmetry which yields the 
desired mixing.  This is well illustrated, for instance, in the 
case of tri-bimaximal mixing for which an extensive literature exists in 
which models based on one of the discrete symmetries mentioned above have 
been proposed.

The plan of the paper is as follows: for each mixing scenario 
considered in Section \ref{sec:schemes}, we have plotted the allowed 
mixing angle ranges and compared them with the present mixing data. 
Conclusions are drawn in Section \ref{sec:concl}. 
Some of the schemes display a high amount of symmetry but require moderate 
perturbations in order to bring them into compliance with the data, and we 
have treated the various possibilities for doing so in several cases in the
Appendix.

\section{\label{sec:schemes}Lepton Mixing Schemes}
We begin with the Pontecorvo-Maki-Nakagawa-Sakata (PMNS)
mixing matrix, which in general is given by 
\be
U = U_\ell^\dagger \, U_\nu \, , 
\ee
where $U_\ell$ ($U_\nu$) stems from diagonalization of the charged lepton
(neutrino) mass matrix. 
The standard form of the PMNS matrix is 
\be \label{eq:U}
U = \left( \bad 
c_{12}  \, c_{13} 
& s_{12} \, c_{13} 
& s_{13} \, e^{-i \delta}  \\ 
-s_{12} \, c_{23} 
- c_{12} \, s_{23} \, 
s_{13}  \, e^{i \delta} 
& c_{12} \, c_{23} - 
s_{12} \, s_{23} \, s_{13} 
\, e^{i \delta} 
& s_{23}  \, c_{13}  \\ 
s_{12}   \, s_{23} - c_{12} 
\, c_{23}  \, s_{13} \, e^{i \delta} & 
- c_{12} \, s_{23} 
- s_{12} \, c_{23} \, 
s_{13} \, e^{i \delta} 
& c_{23}  \, c_{13}  
\ea   
\right) P \,,
\ee
where $c_{ij} = \cos \theta_{ij},\ s_{ij} = 
\sin \theta_{ij}$ with $\delta$ the unknown CP-violating 
Dirac phase. The two equally unknown Majorana phases appear in 
$P = {\rm diag}(1,e^{i\alpha},e^{i \beta})$. 
While the phases are currently unconstrained, 
the present best-fit values of the mixing angles and their 1$\sigma$, 
2$\sigma$ and 3$\sigma$ ranges \cite{Schwetz2010} are presented in 
Table~\ref{tbl:angles} 
(other groups obtain very similar results \cite{other}). 
The above parameterization of $U$ is obtained
by three consecutive rotations:
\bea \label{eq:rot}
U = R_{23}(\theta_{23}) \, \tilde{R}_{13}(\theta_{13}; \delta) \, 
R_{12}(\theta_{12})\, ,  \mbox{where e.g., } \\[0.1in]
R_{12}(\theta_{12}) = 
\left( \bad 
c_{12} & s_{12} & 0 \\
-s_{12} & c_{12} & 0 \\
0 & 0 & 1 
\ea
\right) , \quad
\tilde{R}_{13}(\theta_{13}; \delta)  = 
\left( \bad 
c_{13} & 0 & s_{13} \, e^{-i \delta} \\
0 & 1 & 0 \\ 
-s_{13}  \, e^{i \delta} & 0 & c_{13} 
\ea
\right) .
\eea

\begin{table}[t]
\centering
\begin{tabular}{c|ccc}
\hline\hline \\[0.0in]
\textbf{Parameter}  & $\textbf{Best-fit}^{\boldsymbol{+1 \sigma}}_{\boldsymbol{-1 \sigma}}$ & $\boldsymbol{2 \sigma}$ & 
    $\boldsymbol{3 \sigma}$ \\[0.1in] \hline \\[0.0in]
\textbf{sin}$^\textbf{2} \text{ } \boldsymbol\theta_\textbf{12}$ & 
    0.318$^{+0.019}_{-0.016}$ & 0.29-0.36 & 0.27-0.38 \\
\textbf{sin}$^\textbf{2} \text{ } \boldsymbol\theta_\textbf{23}$ & 
    0.500$^{+0.070}_{-0.060}$ & 0.39-0.63 & 0.36-0.67 \\
\textbf{sin}$^\textbf{2} \text{ } \boldsymbol\theta_\textbf{13}$ & 
    0.013$^{+0.013}_{-0.009}$ & $\le 0.039$ & $\le 0.053$ \\[0.1in] 
\hline \hline 
\end{tabular}
\caption{Mixing angles and their 1$\sigma$, 2$\sigma$ and 3$\sigma$ ranges 
    \cite{Schwetz2010}.}
\label{tbl:angles}
\end{table}

The most popular mixing scenario approximating the current data is the 
tri-bimaximal one \cite{HPS,tbm}: 
\be
U_{\rm TBM} = \left( 
\bad 
\sqrt{\frac 23} & \sqrt{\frac 13} & 0 \\
-\sqrt{\frac 16} & \sqrt{\frac 13} & -\sqrt{\frac 12} \\
-\sqrt{\frac 16} & \sqrt{\frac 13} & \sqrt{\frac 12}
\ea
\right) ,
\ee
corresponding to\footnote{To obtain this form of $U$, it is necessary to
insert $\theta_{23} = - \pi/4$ in the standard parameterization 
(\ref{eq:U}) of the PMNS matrix. Compared to $\theta_{23} = + \pi/4$, the
difference is unphysical, of course. In the following we will use 
$\theta_{23} = - \pi/4$ whenever we speak about maximal atmospheric mixing.}  
\be
\sin^2 \theta_{12} = \frac 13~,~~
\sin^2 \theta_{23} = \frac 12~,~~|U_{e3}| = 0 \,.
\ee
The overwhelming majority of the plethora of models (see 
\cite{AF,jap,BR} for a list of references) invokes the symmetry group
$A_4$. One reason is that $A_4$ is rather economical: it is the smallest 
discrete group containing a three dimensional irreducible representation 
(IR). Furthermore, in the flavor basis it can be generated by two
generators\footnote{Some groups require 3 generators.} 
$S$ and $T$, one of which is diagonal and leaves the charged lepton mass 
matrix diagonal, while the other one leaves $m_\nu^{\rm TBM}$
invariant \cite{AF}, where 
\be \label{eq:mnutbm}
m_\nu^{\rm TBM} = \left( \bad 
A & B & B \\
\cdot & \frac 12 (A + B + D) & \frac 12 (A + B - D) \\
\cdot & \cdot & \frac 12 (A + B - D) 
\ea
\right)
\ee  
is the most general neutrino mass matrix leading to TBM. A geometrical
motivation is provided by noting that $A_4$ is the symmetry group
of the regular tetrahedron, and the angle between two faces is 
$2 \theta_{\rm TBM}$, where $\sin^2 \theta_{\rm TBM} = \frac 13$. 
Models can be constructed in such a way that the Yukawa 
couplings, and hence the mass matrices, are invariant under certain
group elements, which are generated by $S$ and $T$, which in turn are connected
to the symmetry of the geometrical object the group describes. In this
way the connection between geometry and flavor physics can arise.

Tri-bimaximal mixing is a variant of the more general $\mu$--$\tau$ symmetry, 
which leaves solar neutrino mixing unconstrained: 
\be
U_{\mbox{\scriptsize $\mu$--$\tau$}} = 
\left( 
\bad 
\cos \theta_{12} & \sin \theta_{12} & 0 \\
- \frac{\sin \theta_{12}}{\sqrt{2}} &  
 \frac{\cos \theta_{12}}{\sqrt{2}} 
& -\sqrt{\frac 12} \\
- \frac{\sin \theta_{12}}{\sqrt{2}} &  
\frac{\cos \theta_{12}}{\sqrt{2}} 
& \sqrt{\frac 12}
\ea
\right) ,
\ee
corresponding to 
\be
\sin^2 \theta_{23} = \frac 12~,~~|U_{e3}| = 0 \,.
\ee
From a theoretical point of view, $\theta_{12}$ is unconstrained by 
$\mu$--$\tau$ symmetry and hence can be expected to be a number of
order one. This is indeed in good agreement with data. A simple
$Z_2$ or $S_2$ exchange symmetry acting on the neutrino mass matrix
suffices to generate $\mu$--$\tau$ symmetry. In fact, any symmetry
having $Z_2$ or $S_2$ as a subgroup can be used, for instance, 
$D_4$ \cite{D4}.\\

We now turn to other mixing scenarios which serve as alternatives 
to the tri-bimaximal one. 
First consider trimaximal mixing and its variants
\cite{GL0,GL1,lam,AR} (see also \cite{lebed}). 
Here a given row or column of $U$ takes the same form as for 
tri-bimaximal mixing. The term ``trimaximal'' was originally used for
the case of the second column of the PMNS matrix being identical to
the TBM case. The analogous possibilities for the other rows and
columns go under the same banner ``trimaximal''. The notation is such
that if the $i$th column (row) of $U$ has the same form as for TBM, then the
scenario is called TM$_i$ (TM$^i$). 
In case this applies
to the first column of $U$, the condition is: 
\be \label{eq:TMsub1}
\mbox{TM}_1:~~
\left(
\ba 
|U_{e1}|^2 \\
|U_{\mu 1}|^2 \\
|U_{\tau 1}|^2
\ea
\right) = 
\left(
\ba 2/3 \\ 1/6 \\ 1/6
\ea
\right) .
\ee
The implications of this Ansatz are \cite{AR}
\be
\sin^2 \theta_{12} = \frac 13 \, \frac{1 - 3 \, |U_{e3}|^2}
{1 - |U_{e3}|^2} \simeq 
\frac 13 \, \left( 1 - 2 \, |U_{e3}|^2 \right) 
\ee
and 
\be \label{eq:TM_1cond}
\cos \delta \, \tan 2 \theta_{23} = 
 - \frac{1 - 5 \, |U_{e3}|^2}{2\sqrt{2} \, |U_{e3}| \, 
\sqrt{1 - 3 \, |U_{e3}|^2}} \\
\simeq 
\frac{-1}{2\sqrt{2} \, |U_{e3}|} \left(1 - \frac{7}{2} \, 
|U_{e3}|^2 \right) . 
\ee
For the second column the originally-named trimaximal condition is 
\be 
\mbox{TM}_2:~~
\left(
\ba 
|U_{e2}|^2 \\
|U_{\mu 2}|^2 \\
|U_{\tau 2}|^2
\ea
\right) = 
\left(\ba 1/3 \\ 1/3 \\ 1/3 \\ \ea \right)  , 
\ee
leading to \cite{GL0,AR}
\be \label{eq:TM2}
\sin^2 \theta_{12} = \frac 13 \, \frac{1}{1 - |U_{e3}|^2} \ge
\frac 13 
\ee
and 
\bea
\label{eq:TM_2cond} \D
\cos \delta \, \tan 2 \theta_{23} = \frac{2 \, \cos \theta_{13} \, 
\cot 2 \theta_{13}}{\sqrt{2 - 3 \, \sin^2 \theta_{13}}} 
= \frac{1 - 2 \, |U_{e3}|^2}{|U_{e3}| \, 
\sqrt{2 - 3 \, |U_{e3}|^2}}
    \\[0.2in] \D \simeq   
\frac{1}{\sqrt{2}} \frac{1}{|U_{e3}|}
\left(1 - \frac{5}{4}\, |U_{e3}|^2 
\right) .
\eea

If we would insist that the third column of $U_{\rm TBM}$ remains invariant 
instead, i.e., $|U_{e3}|^2 =~0$, $|U_{\mu3}|^2 = |U_{\tau3}|^2 = \frac 12$, 
then $\theta_{13} = 0$, $\theta_{23} = \pi/4$, while $\theta_{12}$ is 
a free parameter and $\delta$ is arbitrary.  This case (${\rm TM_3}$ in our 
notation) is nothing other than $\mu$--$\tau$ symmetry. 

It was argued \cite{lam} that models based on flavor 
symmetries which have $A_4$ as a subgroup should 
be possible for TM$_2$. For TM$_1$ and TM$_3$, these groups are 
$S_4$ and $S_3$, respectively. Models based on flavor symmetry groups 
$\Delta(27)$ \cite{GL0} and $S_3$ \cite{GL1} have been constructed for the 
trimaximal scenario TM$_2$. 

Now consider the case where one of 
the rows of the tri-bimaximal mixing matrix remains invariant \cite{AR}.
We start with the case of the first row in $U_{\rm TBM}$ remaining invariant,
denoting this by TM$^1$, 
\be
\mbox{TM$^1$}:\qquad 
\left( 
|U_{e 1}|^2 \,,~|U_{e 2}|^2\,,~|U_{e 3}|^2 \right) = 
\left( 
\frac 23\,,~\frac 13 \,,~0
\right) .
\ee 
Here $\theta_{23}$ is a free parameter, while 
$\sin^2 \theta_{12} = \frac 13$, as well as $\theta_{13} = \delta =
0$.

If we consider only the second or third row invariant, we can again correlate 
all four mixing parameters. Starting with the second row, i.e., 
\be
\mbox{TM}^2: \qquad 
\left( 
|U_{\mu 1}|^2 \,,~|U_{\mu 2}|^2\,,~|U_{\mu 3}|^2 
\right) = 
\left( 
\frac 16\,,~\frac 13 \,,~\frac 12
\right),
\ee 
one immediately finds from $|U_{\mu 3}|^2 = \frac 12$: 
\be \label{eq:s23v3}
\sin^2 \theta_{23} = \frac{1}{2 \, (1 - |U_{e3}|^2)} 
\simeq \frac 12 \left(1 + |U_{e3}|^2 
\right) \ge \frac 12\,,
\ee
with atmospheric neutrino mixing on the ``dark side'' ($\theta_{23}
\ge \pi/4$). The second correlation among the mixing parameters is 
\be \label{eq:TM^2cond}
\sin^2 \theta_{12} \simeq \frac 13 
- \frac{2\sqrt{2}}{3} \,|U_{e3}|\, 
\cos \delta \,  + \frac 13 \, |U_{e3}|^2 \,\cos 2 \delta \,. 
\ee
On the other hand, with the third row remaining invariant, 
\be
\mbox{TM}^3:~
\left( 
|U_{\tau 1}|^2 \,,~|U_{\tau 2}|^2\,,~|U_{\tau 3}|^2 
\right) = 
\left( 
\frac 16\,,~\frac 13 \,,~\frac 12
\right) ,
\ee 
the atmospheric neutrino mixing is now predicted on the ``bright side,''
($\theta_{23} \le \pi/4$):
\be
\sin^2 \theta_{23} = \frac{1 - 2\, |U_{e3}|^2}{2 \, (1 - 
|U_{e3}|^2)} \simeq \frac 12 \left(1 - |U_{e3}|^2 
\right) \le \frac 12\,,
\ee 
while the solar neutrino mixing is correlated with $|U_{e3}|$ and $\delta$ 
according to 
\be \label{eq:TM^3cond}
\sin^2 \theta_{12} \simeq \frac 13 
+ \frac{2\sqrt{2}}{3} \,|U_{e3}|\, 
\cos \delta \,  + \frac 13 \, |U_{e3}|^2 \,\cos 2 \delta \,. 
\ee

We also note the recently proposed tetramaximal 
mixing scheme (T$^4$M) \cite{tetra}. 
Its name stems from the fact that it can be obtained by four 
consecutive rotations, each having a maximal angle of $\pi/4$, 
and properly chosen phases associated with the rotations: 
\be
U_{\rm tetra} = R_{23} (\pi/4; \pi/2) \, R_{13} (\pi/4; 0) \, 
R_{12} (\pi/4; 0) \, R_{13} (\pi/4; \pi) \,. 
\ee
The notation of the rotation matrices is defined in
Eq.~(\ref{eq:rot}). The definite predictions 
are\footnote{By multiplying a fifth maximal rotation $R_{12}(\pi/4;
2\pi/3)$ to the right of $U_{\rm tetra}$ one could obtain
``quintamaximal mixing'', which has more complicated predictions:  
$\sin^2 \theta_{12} = (3 + \sqrt{2})/(10 + 4 \sqrt{2}) \simeq 0.282$, 
$|U_{e3}|^2 = (3 - 2\sqrt{2})/8 \simeq 0.021$, $\sin^2 \theta_{23} =
\frac 12$ and $J_{\rm CP} = (3\sqrt{2} - 2)/256 \simeq 0.0088$. Here 
$J_{\rm CP} = \text{Im}\{U_{e1} \, U_{\mu2} \, U_{e2}^* \, 
U_{\mu1}^*\}$ is the usual measure for CP violation.} 
\bea
\delta= \pi/2,\quad \sin^2 \theta_{23} = \frac 12~,\quad 
\sin^2 \theta_{12} = (\frac 52 + \sqrt{2})^{-1} \simeq 0.255 \, , 
\\[0.1in]
|U_{e3}|^2 = \frac 14 \, (\frac 74 - \sqrt{2} ) \, \sin^2 \theta_{12}
= \frac 14 \, (\frac 32 - \sqrt{2}) \simeq 0.021 \, .
\eea

Another interesting possible property of $U$ is that it might be
symmetric: $U = U^T$. One can show that there follows one constraint on the 
mixing parameters \cite{HR}: 
\be \label{eq:limit}
|U_{e3}| =  
\frac{\sin \theta_{12} \, \sin \theta_{23}}
{\sqrt{1 - \sin^2 \delta \, \cos^2 \theta_{12} \, \cos^2 \theta_{23}} 
+ \cos \delta \, \cos \theta_{12} \, \cos \theta_{23}}~.
\ee
Scenarios in which this happens at lowest order, for instance, reflect
``Quark-Lepton Universality'' \cite{JS}. Here it is proposed that
down quarks and charged leptons are diagonalized by the same matrix 
$V$ and that the down quark mass matrix is hermitian. 
Furthermore, $m_D = m_{\rm up} = m_{\rm up}^T$, and $M_R$ is also diagonalized
by $V$, where $m_D$ ($M_R$) is the Dirac (Majorana) mass matrix in
the type I seesaw mechanism. With these assumptions it follows that 
the PMNS matrix is symmetric. In general, $U$ is symmetric if 
$U_\ell = S \, U_\nu^\dagger$, where $S$ is a symmetric and unitary
matrix. Moreover, if $m_\nu^\ast$ and (symmetric) $m_\ell$
are diagonalized by the same matrix, again the PMNS matrix is symmetric.\\ 

 Several proposed mixing matrices single out the solar mixing angle 
for special treatment.  In the case of bimaximal mixing (BM), 
$\sin^2 \theta_{12} = 1/2$, with the same atmospheric and reactor neutrino 
mixing angles as in the case of tri-bimaximal mixing or $\mu$--$\tau$
symmetry. Hence the mixing matrix 
has the form \cite{BM}
\be \label{eq:UBM}
U_{\rm BM} = \left( 
\bad 
{\frac 1{\sqrt{2}}} & {\frac 1{\sqrt{2}}} & 0 \\
-{\frac 12} & {\frac 12} & -{\frac 1{\sqrt{2}}} \\
-{\frac 12} & {\frac 12} & {\frac 1{\sqrt{2}}} 
\ea
\right) ,
\ee
In \cite{rabi} it has been shown that for instance 
one can use the discrete symmetry
$S_3$ to construct such a mixing matrix. While the value $\sin^2
\theta_{12} = \frac 12$ is ruled out by close to $10 \sigma$, this
mixing scenario has recently been revived in the form of a model based
on $S_4$ \cite{bm_AFM}. Here the two generators of the group are
chosen such that one is diagonal and the other one leaves $m_\nu^{\rm
BM}$ invariant, where $m_\nu^{\rm BM}$ is the most general mass matrix
leading to bimaximal mixing, which is obtained from
Eq.~(\ref{eq:mnutbm}) by removing $B$. Bimaximal mixing can be
corrected by charged lepton corrections, leading to QLC scenarios 
(see below).

Another possibility proposed here 
is ``hexagonal mixing'' (HM), where $\theta_{12} = \pi/6$, or 
$\sin^2 \theta_{12} = 1/4$. In this case, 
again with maximal atmospheric and vanishing reactor neutrino mixings,
the mixing matrix is given by 
\be
U_{\rm HM} = \left( 
\bad 
{\frac {\sqrt{3}}2} & {\frac 12} & 0 \\
-{\frac 1{2\sqrt{2}}} & {\frac {\sqrt{3}}{2\sqrt{2}}} & -{\frac 1{\sqrt{2}}} \\
-{\frac 1{2\sqrt{2}}} & {\frac {\sqrt{3}}{2\sqrt{2}}} & {\frac 1{\sqrt{2}}} 
\ea
\right) .
\ee
Here $D_{12}$ is an appropriate discrete flavor symmetry. The 
angle $\theta_{12} = \pi/6$ is obviously the external angle of the
dodecagon, whose symmetry group is $D_{12}$. One can also use $D_6$,
where the external angle is $\pi/3$. 
Both BM and HM require corrections to bring them into agreement with
current global fits. A strategy to do this is given in the Appendix. 
Note that this requires a larger correction for bimaximal mixing than for 
hexagonal mixing, where the necessary correction is moderate. 

There are two proposals which link solar neutrino mixing with the
golden ratio angle $\varphi = (1 + \sqrt{5})/2$: 
\begin{eqnarray}
\varphi_1 : & \cot \theta_{12} = &  
\varphi \Rightarrow \sin^2 \theta_{12} 
= \frac{1}{1 + \varphi^2} \simeq 0.276 \, ,\\
\varphi_2 : & \cos \theta_{12} = & 
\frac{\varphi}{2} 
\Rightarrow \sin^2 \theta_{12} 
= \frac 14 \, (3 - \varphi) \simeq 0.345 \, .
\end{eqnarray}
The observation that the first relation is allowed has been made in 
Refs.~\cite{g1}. 
Interestingly, the first relation may be obtained with the choice of 
$A_5$ as the flavor symmetry group, as noted in Ref.~\cite{A5}. 
This follows since $A_5$ is isomorphic to the symmetry group 
of the icosahedron whose 12 vertices separated by edge-length 2 have 
Cartesian coordinates specified by $(0, \pm 1, \pm \varphi)$, 
$(\pm 1, \pm \varphi, 0)$ and $(\pm \varphi, 0, \pm 1)$.  Indeed, one can 
write the generators of one of the three-dimensional IRs of $A_5$ in 
terms of $\varphi$ \cite{A5}. One could in principle assign the values 
$\sin^2 \theta_{23} = \frac 12$ and $U_{e3} = 0$ to the two golden 
ratio relations. 
 
The second golden ratio relation was proposed first in 
\cite{g2}. In Ref.~\cite{ABR} a model based on the discrete flavor 
symmetry $D_{10}$ has been applied to obtain this angle. Believe it or not, 
$\cos \theta_{12} = \frac{\varphi}{2} $ implies nothing other than 
$\theta_{12} = \pi/5$, and therefore arguments similar to those given above 
for hexagonal mixing apply: the angle $\pi/5$ is the external angle 
of a decagon and $D_{10}$ is its rotational symmetry group.\\ 

The final class of alternative mixing scenarios we consider deals with 
Quark-Lepton Complementarity (QLC), which can be used to relate the
quark and lepton mixing matrices. 
The most naive form relates the solar neutrino mixing angle, 
$\theta_{12}$, 
to the quark Cabibbo angle, $\theta^q_{12}$, by \cite{QLC0a,QLC0b}
\be
{\rm QLC}_0 :~\theta_{12} = \frac{\pi}{4} - \theta_{12}^{q} 
\Rightarrow \sin^2 \theta_{12} \simeq 0.280 \, .
\ee
One may assume a similar relation for the 23-sector, 
$\theta_{23} = \frac{\pi}{4} - \theta_{23}^{q}$, leading to 
$\sin^2 \theta_{23} \simeq 0.459$. 

These QLC relations can be approximately obtained by multiplying a
bimaximal matrix, see Eq.~(\ref{eq:UBM}), 
with the CKM (or a CKM-like) matrix. For definiteness,
we stick to the CKM matrix in what follows. It is given in the Wolfenstein
parametrization \cite{Wolf} by 
\be \label{eq:CKM}
V = \left( 
\bad 
1 - \frac{1}{2} \, \lambda^2 & \lambda & A \, \lambda^3 \, 
(\rho - i \eta) \\[0.3cm]
-\lambda & 1 - \frac{1}{2} \, \lambda^2 & A \, \lambda^2 \\[0.3cm]
A \, \lambda^3 \, (1 - \rho + i \eta) & -A \, \lambda^2 & 1 
\ea
\right) + {\cal{O}}(\lambda^4) .
\ee
In analogy to the PMNS matrix it is a product of two unitary 
matrices, $V = V_{\rm up}^\dagger \, V_{\rm down}$, where 
$V_{\rm up}$ ($V_{\rm down}$) is associated with the 
diagonalization of the up- (down-) quark mass matrix. 
As reported in \cite{ckm} the best-fit values and the 
1$\sigma$, 2$\sigma$ and 3$\sigma$ ranges of the parameters 
$\lambda, A,\bar{\rho},\bar{\eta}$ are  
\bea \label{eq:ckm}
\lambda = \sin \theta_C =  
0.2272^{+0.0010, \, 0.0020, \, 0.0030}
_{-0.0010, \, 0.0020, \, 0.0030}~,\\[0.2cm]
A = 0.809^{+0.014, \, 0.029, \, 0.044}
_{-0.014, \, 0.028, \, 0.042}~,\\[0.2cm] 
\bar\rho = 0.197^{+0.026, \, 0.050, \, 0.074}
_{-0.030, \, 0.087, \, 0.133}~,\\[0.2cm] 
\bar\eta = 0.339^{+0.019, \, 0.047, \, 0.075}_{-0.018, \, 0.037, \, 0.057}~, 
\eea  
where 
$\bar\rho = \rho \,(1 - \lambda^2/2)$ and 
$\bar\eta = \eta \,(1 - \lambda^2/2)$. 
From the relation $U = V^\dagger \, U_{\rm BM}$ 
one finds\footnote{Sometimes a 
Georgi-Jarlskog factor of $\frac 13$ appears in model realizations 
of QLC, in which case the results to be presented can be 
obtained approximately by replacing $\lambda$ with 
$\lambda/3$.} 
\bea
{\rm QLC}_1 :~
\sin^2 \theta_{12} \simeq \frac 12 - \frac{\lambda}{\sqrt{2}} 
\, \cos \phi + {\cal{O}}(\lambda^3)~,~~
|U_{e3}| \simeq \frac{\lambda}{\sqrt{2}} + {\cal{O}}(\lambda^3)~,\\[0.1in]
\sin^2 \theta_{23} \simeq \frac 12 - \frac{\lambda^2}{4}(1+4 \, A \, 
\cos (\phi-\omega)) + {\cal{O}}(\lambda^4) \, ,\\
\eea
where $\lambda$ is the sine of the leading 12-entry in $V$, i.e., the sine 
of the Cabibbo angle. The phases $\phi$ and $\omega$ are not related to the 
phase in the CKM matrix but are relative phases \cite{FPR} between 
$U_\ell = V$ and $U_{\rm BM}$, with $\phi$ corresponding to the Dirac phase 
in neutrino oscillations. With the Jarlskog invariant serving as the 
measure of leptonic CP violation, 
\be
J_{\rm CP} = \text{Im}\{U_{e1} \, U_{\mu2} \, U_{e2}^* \, 
U_{\mu1}^*\} \simeq \frac{\lambda}{4\sqrt{2}} \, \sin \phi  
+ {\cal{O}}(\lambda^3) \, , 
\ee 
numerically one finds $|U_{e3}| \simeq 0.160$, 
$\sin^2 \theta_{12} \gs 0.339$, and $|J_{\rm CP}| \ls 0.0274$, 
since $\phi \ls \pi/4.25$ for 
$\sin^2 \theta_{12}$ to be in its allowed 3$\sigma$ range.

To obtain this scenario in a seesaw framework\footnote{Seesaw
realizations of QLC scenarios are studied in detail in \cite{ssQLC}.}, 
an approach somewhat similar to that for 
Quark-Lepton Universality discussed above is possible
\cite{QLC0a,QLC0b}: diagonalization 
of $m_\nu$ is achieved via $m_\nu = U_{\rm BM}^\ast \, m_\nu^{\rm
diag} \, U_{\rm BM}^\dagger$ and produces exact bimaximal mixing. The 
$U_\ell$ matrix diagonalizing the charged lepton mass matrix $m_\ell$  
corresponds to the CKM matrix $V$. With 
$m_\ell = m_{\rm down}^T$, where 
$m_{\rm down}$ is the down-quark mass matrix, it follows that 
the up-quark mass matrix $m_{\rm up}$ is real and diagonal. It is
assumed to correspond to the Dirac mass matrix in the type I seesaw
formula, and this in turn fixes $M_R$. 

Then there is the second QLC scenario, in which the PMNS matrix is given by 
$U_{\rm BM}^\ast\ V^\dagger$. One finds 
\bea
{\rm QLC}_2 :~
\sin^2 \theta_{12} \simeq \frac 12 - \lambda \, \cos \phi + {\cal{O}}
  (\lambda^3)~,~~
|U_{e3}| \simeq \frac{A \, \lambda^2}{\sqrt{2}} + {\cal{O}}(\lambda^3)~,\\
\sin^2 \theta_{23} \simeq \frac 12 + \frac{A \, \lambda^2}{\sqrt{2}} 
  \cos \phi' + {\cal{O}}(\lambda^3)\,,
\eea
where $\lambda$ is the 12-entry, and $A \, \lambda^2$ the 23-entry of 
$V$. Again the phases $\phi$ and $\phi'$ are unrelated to the phase in the
CKM matrix. Note that there is now a correlation between leptonic CP
violation and quark CKM mixing: 
\be
J_{\rm CP} \simeq \frac{A \, \lambda^2}{4\sqrt{2}} \, \sin 
\phi' + {\cal{O}}(\lambda^4) \, .
\ee
Here the type I seesaw realization goes as follows \cite{QLC0b}: 
diagonalization of $m_\nu$ is achieved via 
$m_\nu = U_\nu^\ast \, m_\nu^{\rm diag} \, U_\nu^\dagger$ and $U_\nu$ is 
related to $V$ (in the sense that 
$U_\nu = V^\dagger$). The charged leptons are diagonalized by 
$U_\ell =  U_{\rm BM}^T$. This in turn can be achieved 
when $V_{\rm up} = V^\dagger$, therefore 
$V_{\rm down}$ must be the unit matrix.  With the definition of 
$M_R = V_R^\ast \, M_R^{\rm diag} \, V_R^\dagger$, where 
$V_R = V_{\rm up}^\ast$, we have 
$m_{\rm up} = m_D = V_{\rm up} \, m_{\rm up}^{\rm diag} \, V$, 
and since $V_{\rm up} = V^\dagger$ the neutrino mass matrix   
$m_\nu = -m_D^T \, M_R^{-1} \, m_D$ is diagonalized by the CKM
matrix. Note that QLC1, QLC2 and Quark-Lepton Universality require
that the eigenvalues of the fermion mass matrices differ even
though some of the mixing angles are the same. Such mass matrices may,
e.g., be ``form diagonalizable'' ones \cite{lv}, which means that the
mixing matrix which diagonalizes the mass matrix is 
independent of the values of the
eigenvalues (such as for bimaximal or TBM).\\ 

\begin{table}[t]
\centering
{
\begin{tabular}{l|lr|lr|lr}
\hline\hline 
       &  &  &  &  &  &  \\
\textbf{Scenario} & \multicolumn{2}{c|}{$\boldsymbol{\sin^2 \theta_{12}}$} & 
    \multicolumn{2}{c|}{$\boldsymbol{ \sin^2 \theta_{23}}$} & 
    \multicolumn{2}{c}{$\boldsymbol{ \sin^2 \theta_{13}}$} \\
    & \textbf{min} & \textbf{max} & \textbf{min} & \textbf{max} & 
    \textbf{min} & \textbf{max} \\ \hline 
       &  &  &  &  &  &  \\
\textbf{TBM} & \multicolumn{2}{c|}{0.333} & \multicolumn{2}{c|}{0.500} & 
     \multicolumn{2}{c}{0.000} \\
$\boldsymbol{\mu-\tau}$ & \multicolumn{2}{c|}{$\boldsymbol{-}$} & 
     \multicolumn{2}{c|}{0.500} & \multicolumn{2}{c}{0.000} \\
\textbf{TM$_\textbf{1}$} & 0.296 & 0.333 & \multicolumn{2}{c|}{**} & 
     \multicolumn{2}{c}{$\boldsymbol{-}$}  \\
\textbf{TM$_\textbf{2}$} & 0.333 & 0.352 & \multicolumn{2}{c|}{**} & 
     \multicolumn{2}{c}{$\boldsymbol{-}$}  \\
\textbf{TM$_\textbf{3}$} & \multicolumn{2}{c|}{$\boldsymbol{-}$} & 
     \multicolumn{2}{c|}{0.500} & \multicolumn{2}{c}{0.000} \\
\textbf{TM$^\textbf{1}$} & \multicolumn{2}{c|}{0.333} & 
     \multicolumn{2}{c|}{$\boldsymbol{-}$} & \multicolumn{2}{c}{0.000} \\
\textbf{TM$^\textbf{2}$} & \multicolumn{2}{c|}{**} & 0.500 & 0.528 & 
     \multicolumn{2}{c}{$\boldsymbol{-}$} \\
\textbf{TM$^\textbf{3}$} & \multicolumn{2}{c|}{**} & 0.472 & 
     0.500 & \multicolumn{2}{c}{$\boldsymbol{-}$} \\
\textbf{T$^4$M} & \multicolumn{2}{c|}{0.255} & \multicolumn{2}{c|}{0.500} & 
     \multicolumn{2}{c}{0.021} \\
\textbf{U=U$^\textbf{T}$} & 0.000 & 0.389 & 0.000 & 0.504 & 0.0343 & 0.053 \\
\textbf{BM} & \multicolumn{2}{c|}{0.500} & \multicolumn{2}{c|}{0.500} & 
     \multicolumn{2}{c}{0.000} \\
\textbf{HM} & \multicolumn{2}{c|}{0.250} & \multicolumn{2}{c|}{0.500} & 
     \multicolumn{2}{c}{0.000}\\
\textbf{$\boldsymbol{\varphi_1}$} & \multicolumn{2}{c|}{0.276} & 
     \multicolumn{2}{c|}{0.500} & \multicolumn{2}{c}{0.000} \\
\textbf{$\boldsymbol{\varphi_2}$} & \multicolumn{2}{c|}{0.345} & 
     \multicolumn{2}{c|}{0.500} & \multicolumn{2}{c}{0.000} \\
\textbf{QLC}$_\textbf{0}$ & \multicolumn{2}{c|}{0.280} & 
     \multicolumn{2}{c|}{0.459} & \multicolumn{2}{c}{$\boldsymbol{-}$} \\
\textbf{QLC}$_\textbf{1}$ & 0.331 & 0.670 & 0.442 & 0.534 & 0.023 & 0.029 \\
\textbf{QLC}$_\textbf{2}$ & 0.276 & 0.726 & 0.462 & 0.540 & 0.0005 & 0.0016 
     \\[0.1in] \hline\hline
\end{tabular}
}
\caption{Predictions for $\sin^2 \theta_{12},\ \sin^2 \theta_{23}$, and 
$\sin^2 \theta_{13} = |U_{e3}|^2$ for the different mixing scenarios 
considered.  The appearance of the symbol $\boldsymbol{-}$ indicates a free 
parameter of the model, while the symbol ** indicates a prediction which 
depends upon the unknown $|U_{e3}|$ and phase $\delta$.  The min and max 
values listed are determined from the presently allowed $3\sigma$ range for 
$|U_{e3}|$.} 
\label{tbl:scenarios}
\end{table}%

We summarize the numerical values of all scenarios considered here in
Table \ref{tbl:scenarios}.  For some of the scenarios, all three mixing 
angles are predicted, while in others one or two of the mixing parameters
remain free parameters (indicated by the symbol $\boldsymbol{-}$) 
and are not determined by the models in question.  Aside from the simple
$\mu$--$\tau$ symmetry case also realized with the TM$_3$ scenario, these 
situations arise when the presently 
unknown experimental reactor neutrino angle, $\theta_{13}$, appearing in
the mixing element $|U_{e3}| = \sin \theta_{13}$ remains unpredicted.
Where possible, minimum and maximum values of the mixing parameters are 
determined by adopting the present experimental $3\sigma$ range for the 
mixing element $|U_{e3}|$, and in the cases of the QLC$_1$ and QLC$_2$ 
models, also for the Wolfenstein parameters.  For four of the models, 
one of the mixing angles is constrained by the other mixing parameters, 
cf.~Eqs.~(\ref{eq:TM_1cond}), (\ref{eq:TM_2cond}), (\ref{eq:TM^2cond}), and 
(\ref{eq:TM^3cond}), 
but the actual numerical value relies not only on one knowing $|U_{e3}|$ 
but also the unknown phase $\delta$.  Such constrained predictions are 
indicated by the symbol ** in Table \ref{tbl:scenarios}.

\begin{figure}[ht]
\begin{center}
\epsfig{file=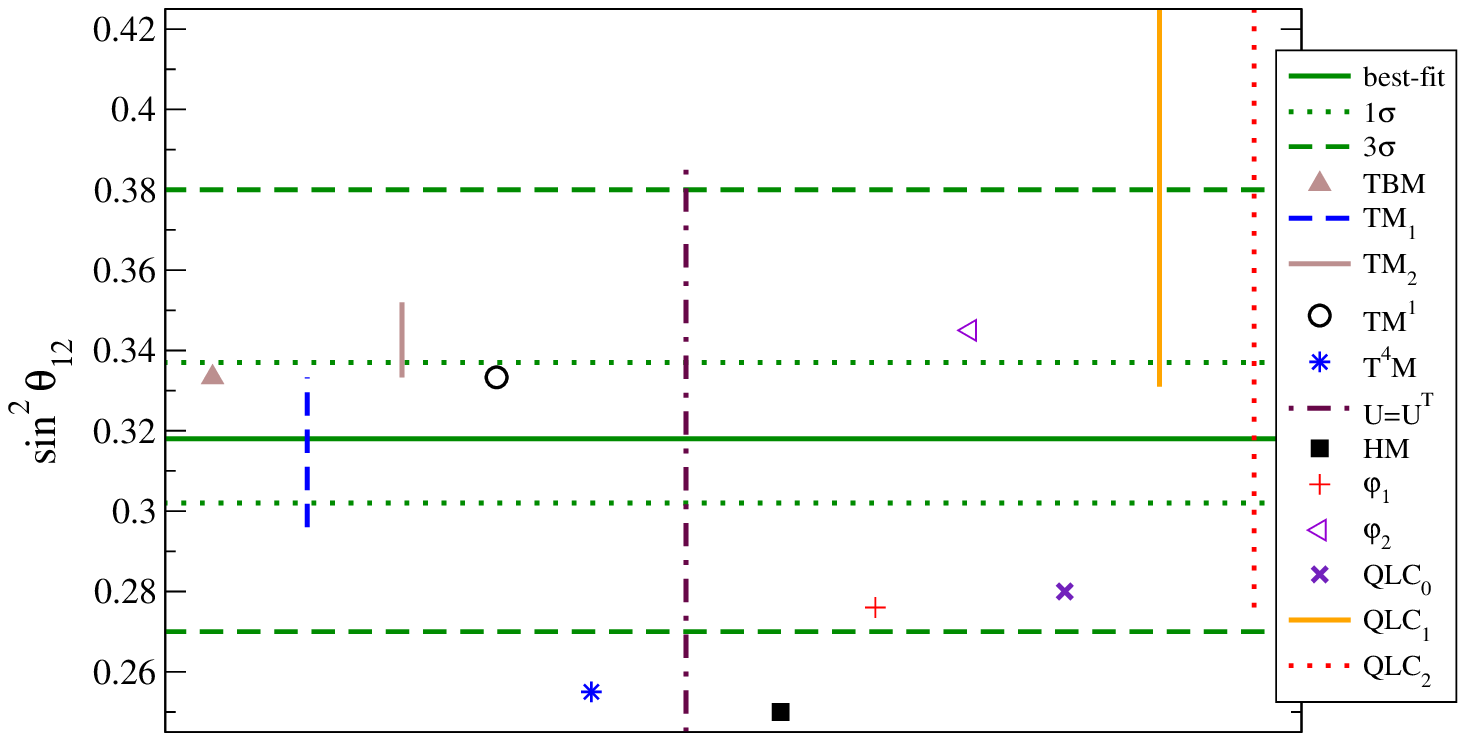,width=14cm,height=9.6cm}
\caption{\label{fig:s12}
Predictions for $\sin^2 \theta_{12}$ of the mixing scenarios 
discussed in the text. For some of the scenarios 
$\sin^2 \theta_{12}$ depends on the other mixing parameters. 
Varying them in their experimentally allowed 3$\sigma$ ranges 
gives the plotted ranges of $\sin^2 \theta_{12}$.}\vspace{.3cm}

\epsfig{file=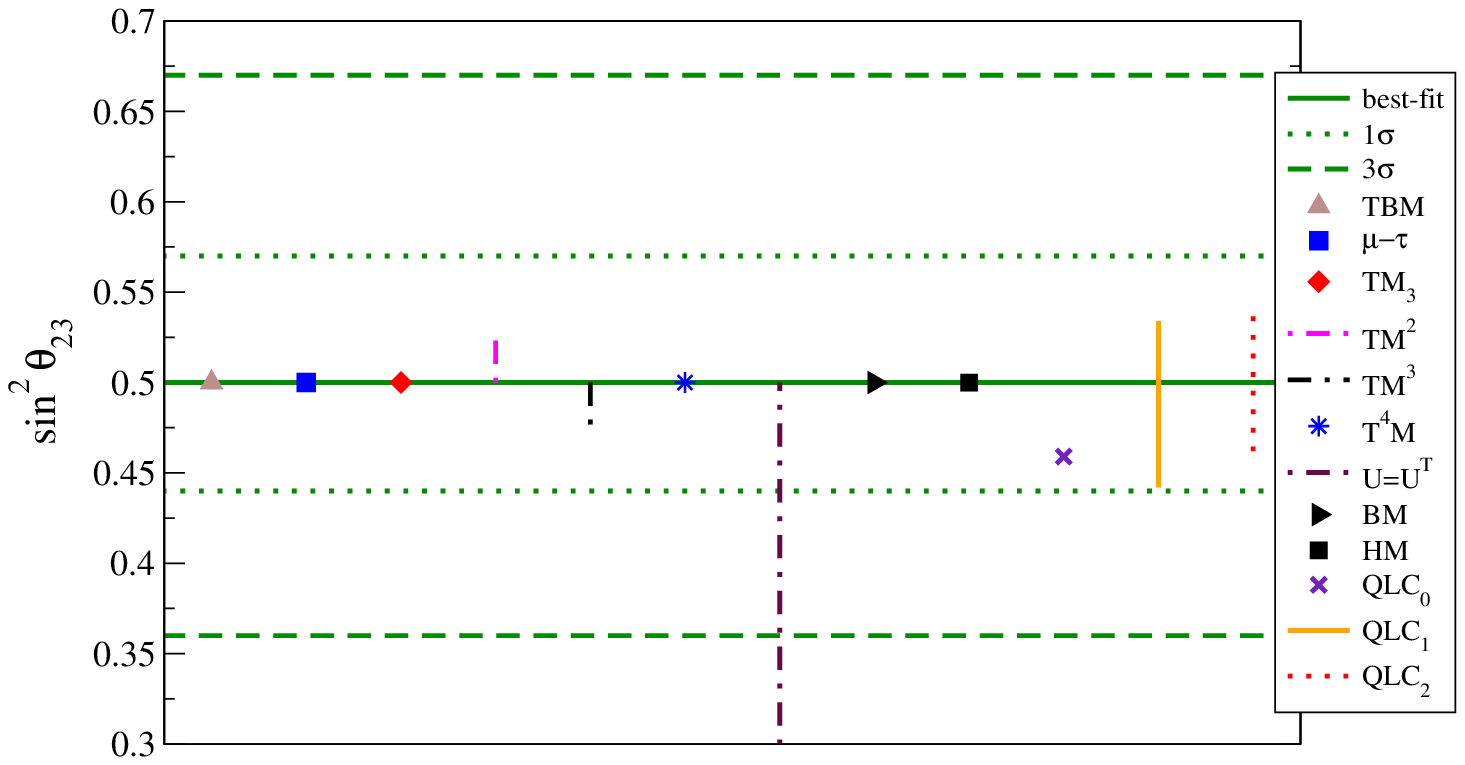,width=14cm,height=9.6cm}
\caption{\label{fig:s23}
Same as Fig.~\ref{fig:s12}, but now for $\sin^2 \theta_{23}$.} 
\end{center}
\end{figure}

\begin{figure}[ht]
\begin{center}
\epsfig{file=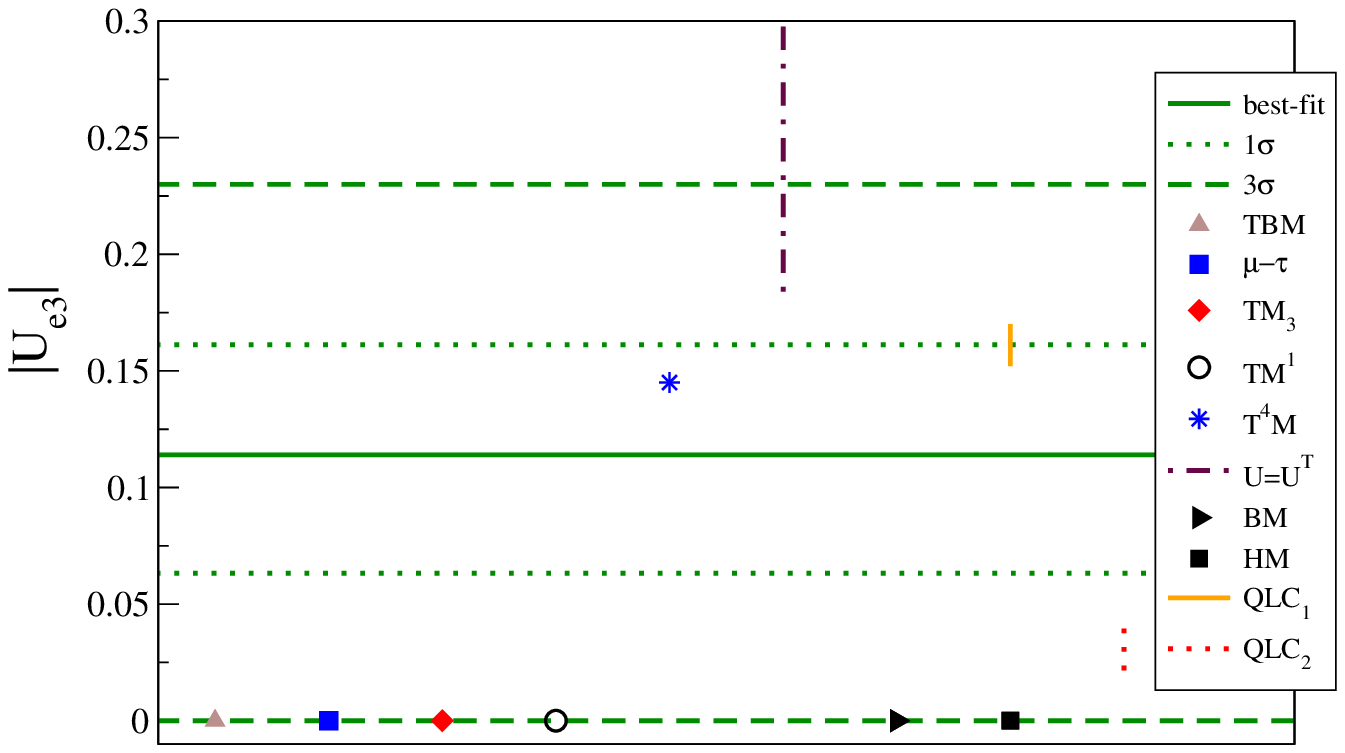,width=13cm,height=9.6cm}
\caption{\label{fig:s13}Same as Fig.~\ref{fig:s12}, but now for 
$|U_{e3}|$.} \vspace{.3cm}
\epsfig{file=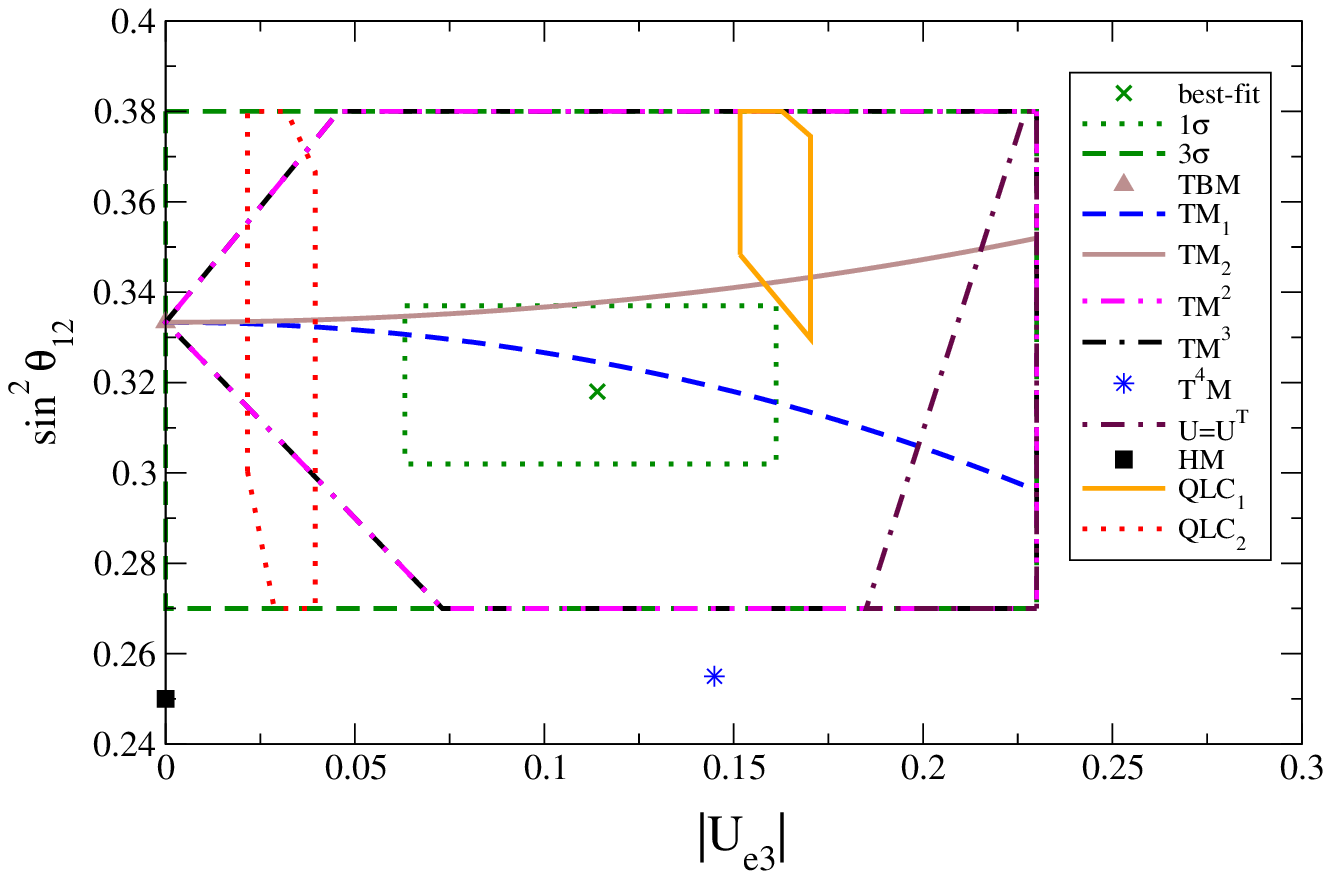,width=13cm,height=9.6cm}
\caption{\label{fig:s12-s13}Correlations between 
$\sin^2 \theta_{12}$ and $|U_{e3}|$ 
constrained by the experimental 3$\sigma$ ranges of the mixing parameters. 
For scenarios where $\sin^2 \theta_{12}$ depends also on the unknown 
Dirac phase $\delta$ the whole area inside the corresponding lines is 
possible, while in the case of TM$_{1,2}$ only parameter combinations 
lying on the dashed (blue) and continuous (brown) lines, respectively, 
are allowed. TM$^2$ and TM$^3$ are here indistinguishable.}

\end{center}
\end{figure}

\begin{figure}[ht]
\begin{center}
\epsfig{file=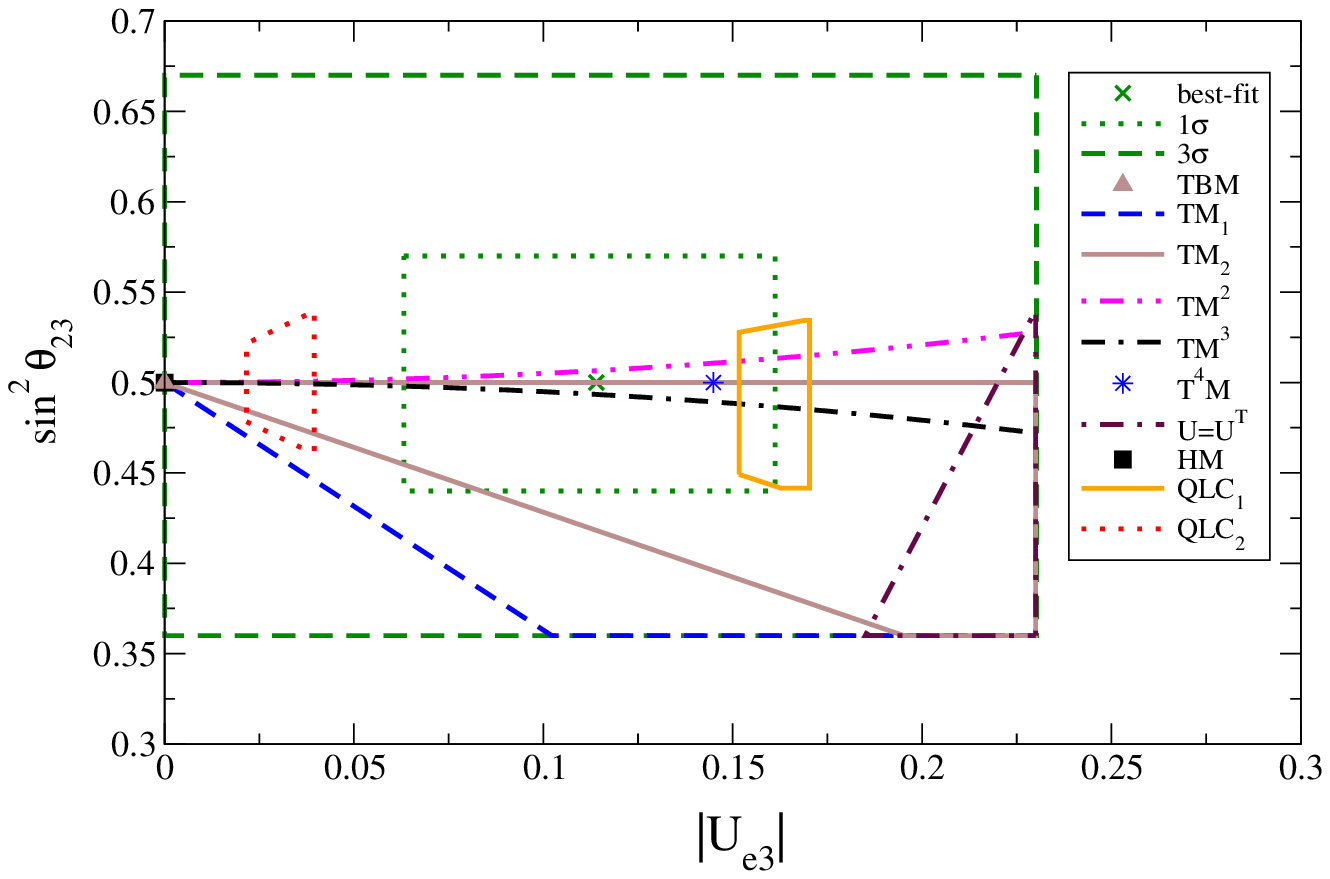,width=13cm,height=9.6cm}
\caption{\label{fig:s23-s13}
Same as Fig.~\ref{fig:s12-s13}, but now for 
$\sin^2 \theta_{23}$. Like $\sin^2 \theta_{12}$ in the TM$_{1,2}$ scenarios, 
in the TM$^{2,3}$ scenarios (magenta and black line, respectively) 
$\sin^2 \theta_{23}$ depends only on $|U_{e3}|$ and not on the 
Dirac phase $\delta$.}\vspace{.3cm}
\epsfig{file=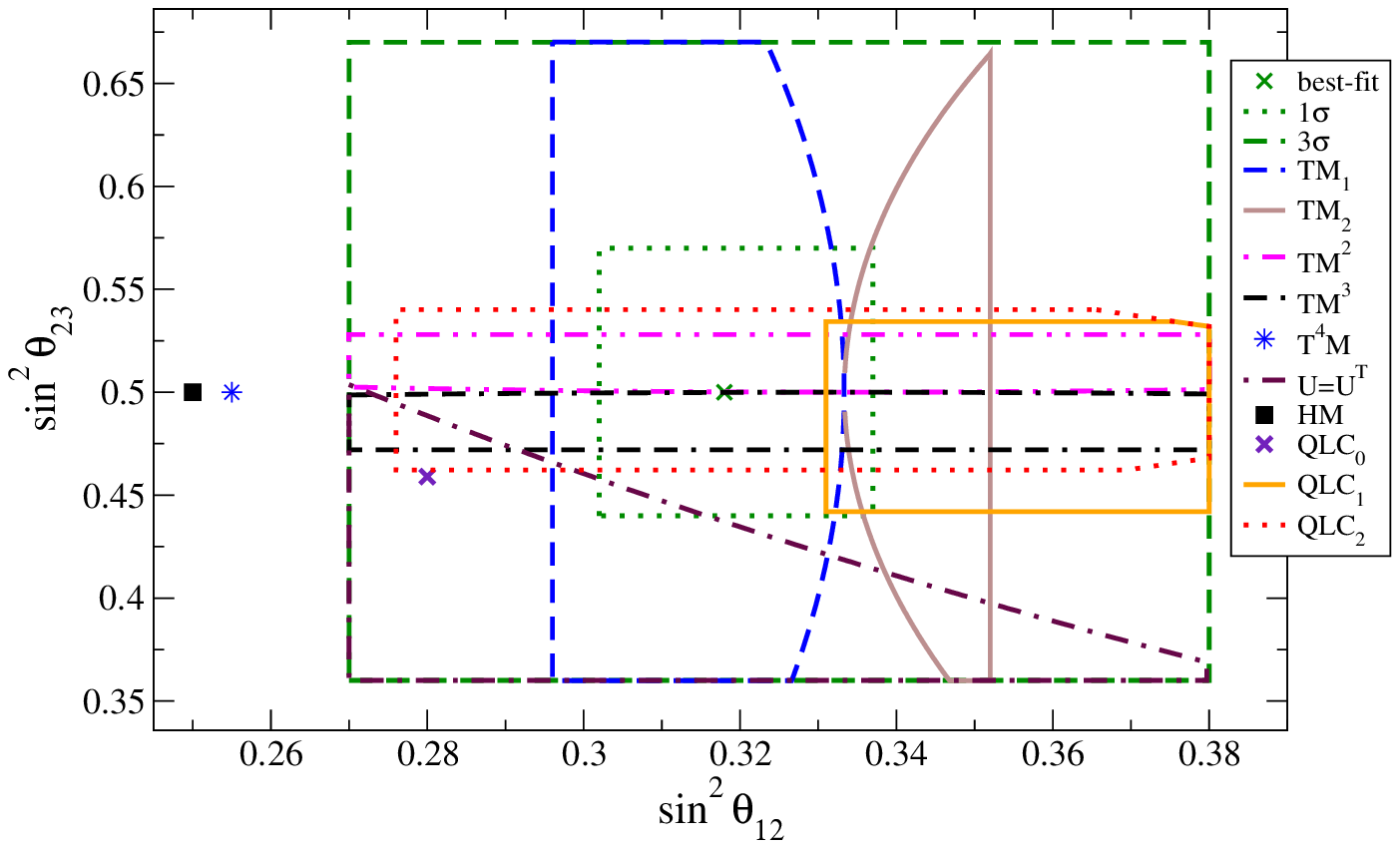,width=14cm,height=9.6cm}
\caption{\label{fig:s23-s12}
Same as Figs. \ref{fig:s12-s13} and \ref{fig:s23-s13}, 
but now the correlations between 
$\sin^2 \theta_{23}$ and $\sin^2 \theta_{12}$ are plotted.}
\end{center}
\end{figure}

In Figs.~\ref{fig:s12}, \ref{fig:s23} and \ref{fig:s13} we plot the 
ranges or values of the three mixing variables, $\sin^2 \theta_{12}$,
$\sin^2 \theta_{23}$, and $|U_{e3}|$, respectively that can be 
obtained for each of the scenarios by varying, if necessary, 
the other variables over their present 
$3\sigma$ experimental range.  The experimentally 
allowed best-fit values, $1\sigma$ and 
$3\sigma$ ranges of the variables are indicated by solid or 
broken horizontal lines as shown in the figures.  
Two-dimensional plots are given in Figs.~\ref{fig:s12-s13}, 
\ref{fig:s23-s13}, and \ref{fig:s23-s12} as functions of $\sin^2 \theta_{12}$
vs.~$|U_{e3}|$, $\sin^2 \theta_{23}$ vs.~$|U_{e3}|$, and $\sin^2 \theta_{23}$
vs.~$\sin^2 \theta_{12}$, respectively. The correlations 
between those observables can be crucial to distinguish scenarios with similar 
predictions.

It is clear from the figures that most of the models cover the 
presently allowed ranges of the mixing angles, with the notable exceptions
of the bimaximal and hexagonal mixing models, BM and HM.  For these models,
one needs to make perturbations on the zeroth order results given in 
Table \ref{tbl:scenarios}. 
We present in Appendix A a simple procedure to perturb the 
hexagonal and bimaximal mixing matrices, as well as the relevant procedure
for the quark-lepton complementarity models, in order to bring their 
results into better agreement with the data.

\section{\label{sec:concl}Conclusions}

With more refined neutrino mixing data available, it is clear that TBM 
gives a reasonably accurate lowest order approximation to the PMNS
mixing matrix.  With this in mind, many authors have constructed top-down 
models based on some discrete flavor symmetry group which yield TBM mixing 
as a natural consequence.  Of the possible choices, the $A_4$ group appears 
to be the most favored choice based on its simplicity.  

We have argued in this 
paper, however, that other possible approximations to the mixing matrix 
exist such as trimaximal mixing or its variants, tetramaximal mixing, 
a symmetric mixing matrix, bimaximal and hexagonal mixings, and mixings 
based on the golden ratio angle or quark-lepton complementarity.  Many of 
these scenarios have already been discussed in the literature, but we have 
compiled this list in order to make easy comparisons of their predictions.
For those requiring perturbations to bring them into better agreement with
the data, we have illustrated how triminimal perturbations of the bimaximal,
and hexagonal mixings or quark-lepton complementarity, for example, can 
accomplish this.  For each one of the starting mixing matrix assumptions, 
one can then use a bottom-up approach to determine the appropriate neutrino
mass matrix from which a suitable discrete flavor symmetry will presumably
reproduce the observed mixing matrix.

The theoretical literature focusses heavily on TBM, and it would be dangerous 
to avoid looking for and studying alternatives. 
We hope that the present paper contributes to the required attention 
on alternatives.

\vspace{0.3cm}
\begin{center}
{\bf Acknowledgments}
\end{center}
We thank Alexei Smirnov for comments on the manuscript. 
This work was supported by the ERC under the Starting Grant 
MANITOP and by the DFG in the Transregio 27 (A.D.~and W.R.). 
One of us (C.H.A.) thanks the Fermilab Theoretical Physics Department,
where part of this work was carried out, for its kind hospitality.  
Work supported by the U.S.~Department of Energy under contract 
No.~DE-AC02-07CH11359.

\renewcommand{\theequation}{A\arabic{equation}}
\setcounter{equation}{0}
\renewcommand{\thetable}{A\arabic{table}}
\setcounter{table}{0}

\begin{appendix}
\section{Perturbing Hexagonal and Bimaximal Mixing}
We begin by discussing the hexagonal mixing Ansatz, defined by 
\be \label{eq:main}
\theta_{12}^\ell = \frac{\pi}{6} = 30^\circ \Rightarrow 
\sin^2 \theta_{12}^\ell = \frac 14 \, , 
\ee
together with maximal $\theta_{23}^\ell$ and $\theta_{13}^\ell
= 0$. 
From now on we denote lepton (quark) mixing angles with a superscript $\ell$
($q$). For this scenario the unperturbed mixing matrix in the lepton mass 
basis reads 
\be
U_{\rm HM} = \left( 
\bad 
\sqrt{\frac 34} & \frac 12 & 0 \\ 
-\frac{1}{2\sqrt{2}} & \sqrt{\frac 38} & -\sqrt{\frac 12} \\
-\frac{1}{2\sqrt{2}} & \sqrt{\frac 38} & \sqrt{\frac 12}
\ea
\right) P \, ,
\ee
where the Majorana phases are contained in 
$P = {\rm diag}(1,e^{i \alpha}, e^{i \beta})$. 
The mass matrix in the charged lepton basis is given by 
$m^0_\nu = U^\ast \, m_\nu^{\rm diag} \, U^\dagger $ and has the texture 
\be \label{eq:mnu}
m^0_\nu = 
\left(
\bad 
A & B & B \\[0.2cm]
\cdot & \frac{1}{2} (A + \sqrt{\frac 83} \, B + D) 
& \frac{1}{2} (A +  \sqrt{\frac 83} \, B - D)\\[0.2cm]
\cdot & \cdot & \frac{1}{2} (A +  \sqrt{\frac 83} \, B + D)
\ea 
\right)\,,
\ee
where the masses and Majorana phases are contained in 
\be
A - \sqrt{\frac 23} \, B = m_1~,~~
A + \sqrt{6} \, B = m_2 \, e^{-2i \alpha}~,~~D = m_3 \, e^{-2i\beta}\,.
\ee
We can also write 
\bea \D \nonumber 
m_\nu^0 = \frac{m_1}{4} 
\left( 
\bad 
3 & -\sqrt{\frac 32} & -\sqrt{\frac 32} \\
\cdot & \frac 12 & \frac 12 \\ 
\cdot & \cdot & \frac 12
\ea
\right) + \frac{m_2 \, e^{-2i \alpha}}{4} 
\left( 
\bad 
1 & \sqrt{\frac 32} & \sqrt{\frac 32} \\
\cdot & \frac 32 & \frac 32 \\ 
\cdot & \cdot & \frac 32
\ea
\right) + \frac{m_3 \, e^{-2i \beta}}{2} 
\left( 
\bad 
0 & 0 & 0 \\
\cdot & 1 & -1 \\ 
\cdot & \cdot & 1 
\ea
\right) \\ \D \\[0.1in]
= m_1 \,\Phi_1 \, \Phi_1^T +  m_2 \, e^{-2i \alpha} \,
\Phi_2 \, \Phi_2^T + m_3 \, e^{-2i \beta} \, \Phi_3 \, \Phi_3^T
\, ,
\eea
where $\Phi_{1,2,3}$ are the columns of the mixing matrix. 
In this limit the $ee$ element of $m_\nu^0$, whose magnitude governs the
rate of neutrino-less double beta decay vanishes when the 
Majorana phase is such that $e^{-2i\alpha} = -1$ 
and in addition the relation $m_1 = \frac 13 \, m_2$, 
or $m_1^2 = \dms/8$ holds.

Independent on the source of perturbation, the most general way to
describe deviations from hexagonal mixing is \cite{PRW} (see also 
\cite{PRW_alt})
\be \label{eq:PRW}
U = R_{23}(-\pi/4) \, U_\epsilon \, R_{12}(\pi/6) \, ,\mbox{ where } 
 U_\epsilon = R_{23}(\epsilon_{23}^\ell) \, 
\tilde{R}_{13}(\epsilon_{13}^\ell; \delta^\ell) 
\, R_{12}(\epsilon_{12}^\ell)\,.
\ee
Note that the order of the small rotations in $U_\epsilon$ is chosen
such that it corresponds to the order of rotations in the usual
description of a mixing matrix. This ``triminimal'' \cite{PRW} 
parametrization implies that each small parameter is 
responsible for only one observable\footnote{A similar strategy may 
be applied to tetra-maximal mixing, where $\theta_{12}$ lies slightly
below the current $3\sigma$ range.}
. The observables are obtained from 
Eq.~(\ref{eq:PRW}) as follows: 
\bea \label{eq:a6}
\sin^2 \theta_{12}^\ell = \frac 14 \left(\cos \epsilon_{12}^\ell 
+ \sqrt{3} \, \sin \epsilon_{12}^\ell \right)^2 \simeq \frac 14 
\left(1 + 2 \sqrt{3} \, \epsilon_{12}^\ell + 3 \, (\epsilon_{12}^\ell)^2
\right) ,\\[0.1in]
\sin^2 \theta_{23}^\ell = \frac 12 - \cos \epsilon_{23}^\ell \, 
\sin \epsilon_{23}^\ell \simeq  \frac 12 - \epsilon_{23}^\ell \, ,\\[0.1in]
U_{e3} = \sin \epsilon_{13}^\ell \, e^{-i\delta^\ell} \, .
\eea
Note that $U_{e3}$ agrees with its form in the usual parameterization 
and that the deviation from maximal atmospheric mixing is to very good
precision given by $\epsilon_{23}^\ell$. 
Regarding solar neutrino mixing, the values $\sin^2 \theta_{12}^\ell$ 
of 0.318, 0.302, 0.337, 0.27, 0.38 and $\frac 13$ are obtained for 
$\epsilon_{12}^\ell = $ 0.076, 0.058, 0.096, 0.023, 0.141, and 0.092. 

In the same way we can perturb the bimaximal mixing matrix, given by 
Eq.~(\ref{eq:UBM}). 
The triminimally perturbed bimaximal mixing matrix can be written as
\be \label{eq:UBMpert}
U = R_{23}(-\pi/4) \, U_\epsilon \, R_{12}(\pi/4) \, ,
\ee
with $U_\epsilon$ the same as in Eq.~\eqref{eq:PRW}. The observables
are obtained as 
\bea \label{eq:UBMobs}
\sin^2 \theta_{12}^\ell = \left(\frac 12 + \sin \epsilon_{12}^\ell
\cos \epsilon_{12}^\ell \right) \simeq \frac 12 + \epsilon_{12}^\ell
\, ,\\[0.1in]
\sin^2 \theta_{23}^\ell = \left(\frac 12 - \sin \epsilon_{23}^\ell
\cos \epsilon_{23}^\ell \right) \simeq \frac 12 - \epsilon_{23}^\ell \,,
\\[0.1in]
U_{e3} = \sin \epsilon_{13}^\ell \, e^{-i\delta^\ell} \, .\\
\eea
Compared to the hexagonal mixing scenario, the values $\sin^2 \theta_{12}^\ell$ 
of 0.318, 0.302, 0.337, 0.27, 0.38 and $\frac 13$ are obtained for 
$\epsilon_{12}^\ell =  -0.186, -0.204, -0.166, -0.239, -0.121$, and $-0.170$.

Returning to hexagonal mixing, one may discuss a related parametrization for
the CKM matrix in the spirit of QLC. 
Namely, with the requirement that the 12-mixing angles
of the quark and lepton sector add up to 45 degrees, it follows
automatically that 
\be
(\theta_{12}^q)^0 = 15^\circ = \frac{\pi}{12} 
\Rightarrow \sin (\theta_{12}^q)^0 
= \frac{\sqrt{3} - 1}{2\sqrt{2}} = 0.2588\,,
\ee
Note that at zeroth order $\theta_{12}^\ell = 2 \theta_{12}^q $. 
There are models in the literature leading to this angle
$(\theta_{12}^q)^0$ \cite{15}. 
In the spirit of triminimality, we can describe the 
necessary but small deviations of this scheme with 
\be
V = R_{23} (\epsilon_{23}^q) \, \tilde{R}_{13}
(\epsilon_{13}^q; \delta^q) \, R_{12} (\epsilon_{12}^q) \, 
R_{12} (\pi/12)\,.
\ee
The sine of the 12-mixing angle is given by 
\be
\sin \theta_{12}^q = \frac 12 \, 
\sqrt{2 - \sqrt{3} \, \cos 2 \epsilon_{12}^q + \sin 2 \epsilon_{12}^q}
\simeq \frac{\sqrt{3} - 1}{2\sqrt{2}}  
\left(
1 + (2 + \sqrt{3}) \, \epsilon_{12}^q
\right) .
\ee
Note that the last expression is equivalent to 
$\sin \theta_{12}^q \simeq \sin (\theta_{12}^q)^0 + \epsilon_{12}^q \,
\cos (\theta_{12}^q)^0$. 
Numerically we have $\sin \theta_{12}^q \simeq 0.2588 + 0.9659 \, 
\epsilon_{12}^q$, so that $\epsilon_{12}^q$ can be almost directly 
identified with the deviation of the sine of Cabibbo angle from 
$\frac{\sqrt{3} - 1}{2\sqrt{2}}$. In order to bring 
$\sin \theta_{12}^q$ into the observed 1$\sigma$ or 3$\sigma$ range given in 
Eq.~(\ref{eq:ckm}) one requires 
\be \label{eq:eps12q}
\epsilon_{12}^q = -0.0326^{+ 0.00102,0.00308}_{- 0.00102,0.00308} \,.
\ee
Note that here $\epsilon_{12}^q$ is negative, while $\epsilon_{12}^\ell$ 
(see Eq.~(\ref{eq:a6}))
is positive. Choosing the tempting value $\epsilon_{12}^\ell = - 
\epsilon_{12}^q$ gives $\sin^2 \theta_{12}^\ell \simeq 0.279$. 

We finish by noting an interesting observation made in 
Ref.~\cite{sv}: taking the
golden ratio relation $\varphi_1$ ($\tan \theta_{12}^\ell = 1/\varphi$)
at face value, and assuming QLC ($\theta_{12}^\ell + \theta_{12}^{q} = \pi/4$) gives 
\be
\tan \theta_{12}^q = \tan (\pi/4 - \theta_{12}^\ell) 
= \frac{1 - 1/\varphi}{1 + 1/\varphi} 
= \frac{1}{\varphi^3}\, , 
\ee
or $\sin \theta_{12}^q \simeq 0.2298$. Hence, the golden ratio may
appear in the quark sector as well. 

\end{appendix}

\vspace*{1in}

\pagestyle{empty}

\end{document}